\documentclass[12pt]{iopart}
\usepackage{graphicx}

\begin{document}
\title[The emergence of cooperativity accompanying vitrification]
{The emergence of cooperativity accompanying vitrification: Insights from density fluctuation dynamics }
\author{Akira Furukawa \footnote{Correspondence author: furu@iis.u-tokyo.ac.jp} } 
\address{Institute of Industrial Science, University of Tokyo, Meguro-ku, Tokyo 153-8505, Japan }
\ead{furu@iis.u-tokyo.ac.jp}
\vspace{10pt}
\begin{indented}
\item[]January 2019
\end{indented}

\begin{abstract}
  We discuss the emergence and growth of the cooperativity accompanying vitrification based on the density fluctuation dynamics for fragile glass-forming liquids. (i) The relaxation of density fluctuations proceeds by the particle (density) exchange process, and is diffusive; that is, the allowable kinetic paths are strongly restricted by the local conservation law. (ii) In normal liquid states, this exchange process is less cooperative, and the diffusion coefficient of density fluctuations $D_c$ is given as $D_c\sim \lambda^2/\tau_\alpha$, where $\lambda$ is the particle size and $\tau_\alpha$ is the structural relaxation time. On the other hand, in supercooled states the restriction on the kinetic path is more severe with increasing the degree of supercooling, which makes the exchange process more cooperative, resulting in $D_c \sim \xi_{\rm d}^2/\tau_\alpha$ with $\xi_{\rm d}$ being the cooperative length scale. (iii) The molecular dynamics simulation results show that the self-diffusion coefficient of the tagged particle, $D_s$, almost coincides with $D_c$, suggesting that the collective density diffusion and the particle diffusion closely share the same mechanism:  in normal states $D_s$ determines $D_c$, but vice versa in supercooled states.  This immediately leads to the idea that the breakdown of the Stokes-Einstein relation is not the anomaly in the single-particle dynamics but reflects the increase in the cooperativity in the density diffusion at the length scale of $\xi_{\rm d}$. 
\end{abstract}

\section{Introduction}

When a liquid is supercooled and approaches the glass transition point, the viscosity $\eta$, which characterizes the flow resistance,  dramatically increases. This drastic slow down behavior defines the glass transition, whose mechanism and origin is still debated today \cite{Ediger-Angell-Nagel,Sillescu,Ediger,Debenedetti-Stillinger,Binder-KobB,DyreR,GotzeB,Berchier-BiroliR,dynamic_heterogeneityB,Tanaka_review}. One of the most striking properties of glass is that it has a liquid-like random structure but shows a solid-like rigidity even without any long range crystalline structures. Certainly, when looking at the static density patterns of glasses, it is difficult to distinguish them from those of ordinary liquids. In other words, the two-body density correlator hardly shows any anomalous or long range features despite the vast differences in the dynamic properties between liquids and glasses. This almost invariant property of the correlation of density fluctuations during the vitrification is in strong contrast to the vastly increasing viscosity. Therefore, in glass physics it has been generally thought that density fluctuations cannot capture any significant signs of the vitrification, and recently (hidden) thermodynamic anomalies and the associated growing static structures have been invoked as the origin of the vitrification \cite{Tanaka_review,Kirkpatrick-Thirumalai-Wolynes,Lubchenko-Wolyness,Kivelson-Kivelson-Zhao-Nussinov-Tarjus}.  Contrary to the conventional or majority views of the glass transition, in a series of recent papers the present author continued to argue that the local density fluctuations themselves strongly control the vitrification \cite{FurukawaG1,FurukawaG2,FurukawaG3,FurukawaG4,FurukawaG5,FurukawaG6}. In this paper, for fragile glass-forming liquids, we specifically discuss how the emergence and growth of the cooperativity associated with vitrification can be captured by the two body {\it dynamic} density correlator additionally showing the supporting molecular dynamics simulation results. 
Here we emphasize that, so far, very little attention has been paid to the potential of the length-scale dependent hydrodynamic transport of density fluctuations to investigate the cooperative effects in glass-forming liquids. Although some parts of the arguments and analysis provided below were already presented in Refs. \cite{FurukawaG1,FurukawaG2,FurukawaG3,FurukawaG4,FurukawaG5,FurukawaG6}, here we show new evidence that the mutual density exchange, which is an elementary process of the density diffusion, is more cooperative with increasing the degree of supercooling. This finding bridges the gap between the observed nonlocal density diffusion and the previously proposed phenomenology for it. Furthermore, we argue that the breakdown of the Stokes-Einstein relation \cite{Fujara,Cicerone-Ediger,Angell-Ngai-McKenna-Mcmillan-Martin} can be attributed to this cooperative exchange dynamics. In this sense, this breakdown is not an anomaly in the single particle dynamics but simply reflects the cooperative effects.   

\section{Density fluctuation dynamics}
\subsection{Collective density diffusion} 

First, let us start from the wavenumber ($k$) dependent relaxation time of the (number) density fluctuations, $\tau_n(k)$. In Fig. 1, we show $\tau_n(k)$ at various temperatures $T$ for two typical models of fragile glass-formers; the Kob-Andersen (KA) \cite{Kob-Andersen} and the Bernu-Hiwatari-Hansen (BHH) soft-sphere \cite{Bernu-Hiwatari-Hansen} models. Please refer to the Appendix for details on the simulation models. We determine $\tau_n(k)$ by fitting a stretched exponential function $A_k \exp\{-[t/\tau_n(k)]^{\beta_k}\}$, where $A_k$ and $\beta_k$ are the $k$-dependent coefficient and exponent, respectively, to the long-time decay of the scaled autocorrelation function, $G(k,t)={\langle n_{{\mbox{\boldmath$k$}}}(t)n_{{-\mbox{\boldmath$k$}}}(0)\rangle}/{\langle|n_{{\mbox{\boldmath$k$}}}|^2\rangle}$.
Here, $n_{{\mbox{\boldmath$k$}}}=\sum_i^N e^{-i{\mbox{\boldmath$k$}}\cdot{\mbox{\boldmath$r$}}_i}$ is the Fourier space representation of the particle number density and $\mbox{\boldmath$r$}_i$ is the position vector of the $i$-th particle. Hereafter, $\langle \cdots\rangle$ denotes the ensemble average and $f_{\mbox{\boldmath$k$}}$ is defined as the Fourier transform of an arbitrary field variable $f({\mbox{\boldmath$r$}})$. The relaxation dynamics exhibit a diffusive nature at longer length scales, satisfying $\tau_n(k)\sim k^{-2}$ \cite{FurukawaG1,FurukawaG2,FurukawaG3}. As was already discussed in Ref. \cite{FurukawaG2}, this diffusive relaxation of density fluctuations can be attributed to local density conservation: in fragile glass formers, the relaxation of density fluctuations proceeds by the local particle exchange process \cite{FurukawaG3}. That is, the increase or decrease in the surrounding particles is accompanied by mutual density exchanges between the nearest neighbors, which is the elementary process of the density diffusion. We insist that such a slow diffusive density relaxation itself is anomalous and cannot be explained by conventional hydrodynamic theory, which distinguishes (fragile) glass forming liquids from ordinary simple liquids. This point will be discussed below. It is noteworthy that such exchange processes are almost absent in liquid silica \cite{FurukawaG3,FurukawaG6}, which is a prototypical strong glass former. In liquid silica, for length scales of several tens of particle sizes (corresponding to approximately 100 \AA), density fluctuations can relax without such a mutual density exchange, and $\tau_n(k)$ exhibits a nearly flat $k$ dependence, which translates to the nondiffusive decay of density fluctuations. This is in marked contrast to the density relaxation in fragile glass formers, suggesting that the role of density fluctuations in the vitrification is quite different between strong and fragile glass formers.  

\begin{figure}[htb]
\includegraphics[width=.9\textwidth]{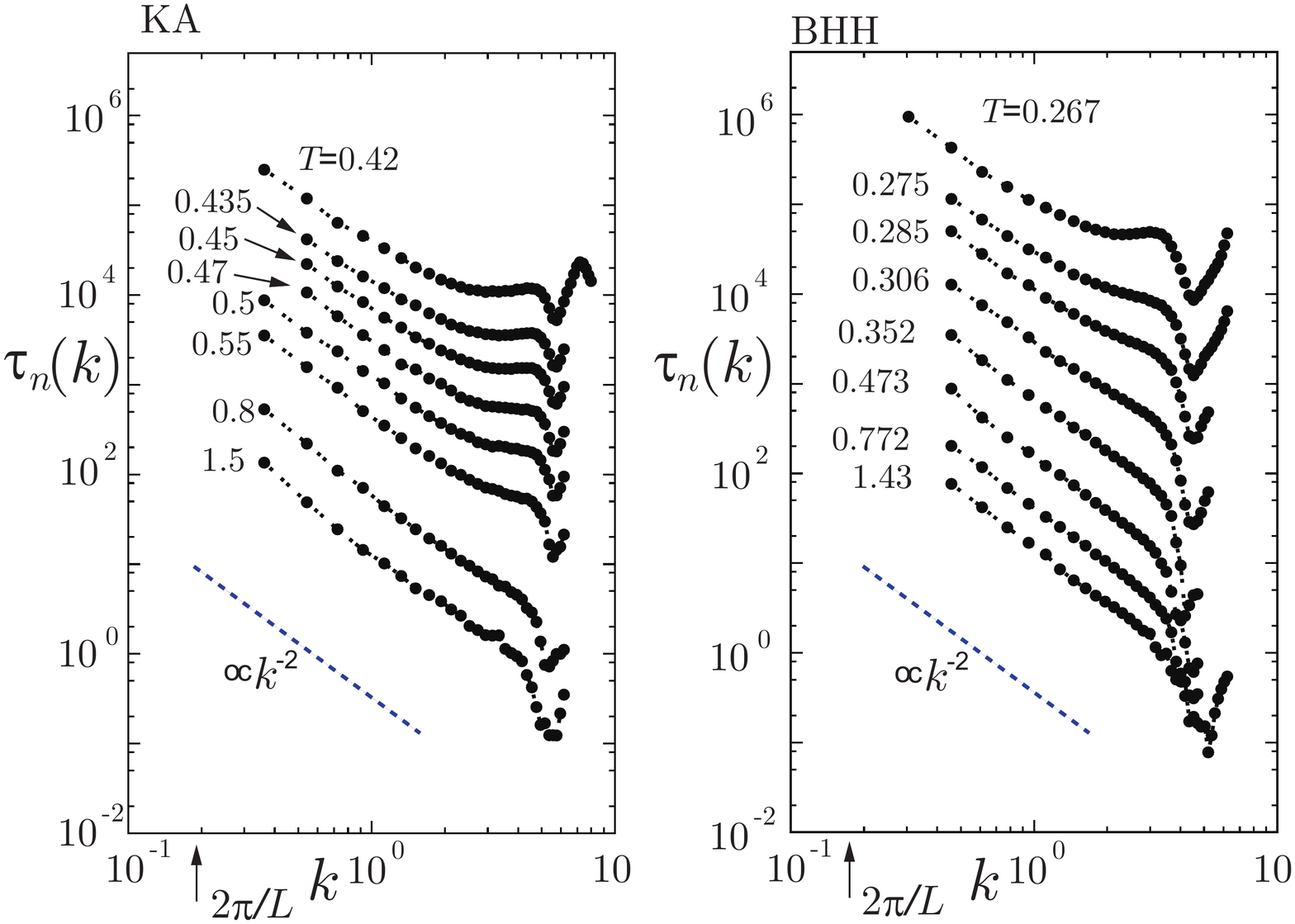}
\caption{ The $k$-dependent relaxation time of number density fluctuations $\tau_n(k)$ at various temperatures for the KA and BHH models. The relaxation dynamics is diffusive at longer length scales as $\tau_n(k)\sim k^{-2}$. Although these plots are almost the same as those shown in Fig. 2 of Ref. \cite{FurukawaG3}, the data were updated.  
}
\label{Fig1}
\end{figure}

In higher temperature normal states, the diffusive behavior already begins at a rather high $k$. However, with increasing the degree of supercooling (decreasing $T$), a deviation from the diffusive behavior is significant for smaller $k$ (larger length scales), which is evident in Fig. 2, where the $k$-dependent density-diffusion coefficient $D_n(k)$ is shown. Here, $D_n(k)$ is defined by $D_n(k)=1/[k^2\tau_n(k)]$. The macroscopic density-diffusion coefficient is formally identified as $D_c=\lim_{k\rightarrow 0} D_n(k)$. In Refs. \cite{FurukawaG1,FurukawaG2,FurukawaG3}, we have interpreted this increasing deviation itself as a reflection of the growth of the cooperativity. In supercooled states, there is a dynamical correlation length of the density diffusion $\xi_{\rm d}$: the relaxation of larger scale ($> \xi_{\rm d}$) fluctuations exhibits diffusive decay, where $\xi_{\rm d}$ can be regarded as a unit size, while smaller scale ($< \xi_{\rm d}$) fluctuations are subordinate to the collective dynamics for the duration of the structural relaxation time $\tau_\alpha$.
\begin{figure}[bth] 
\includegraphics[width=.9\textwidth]{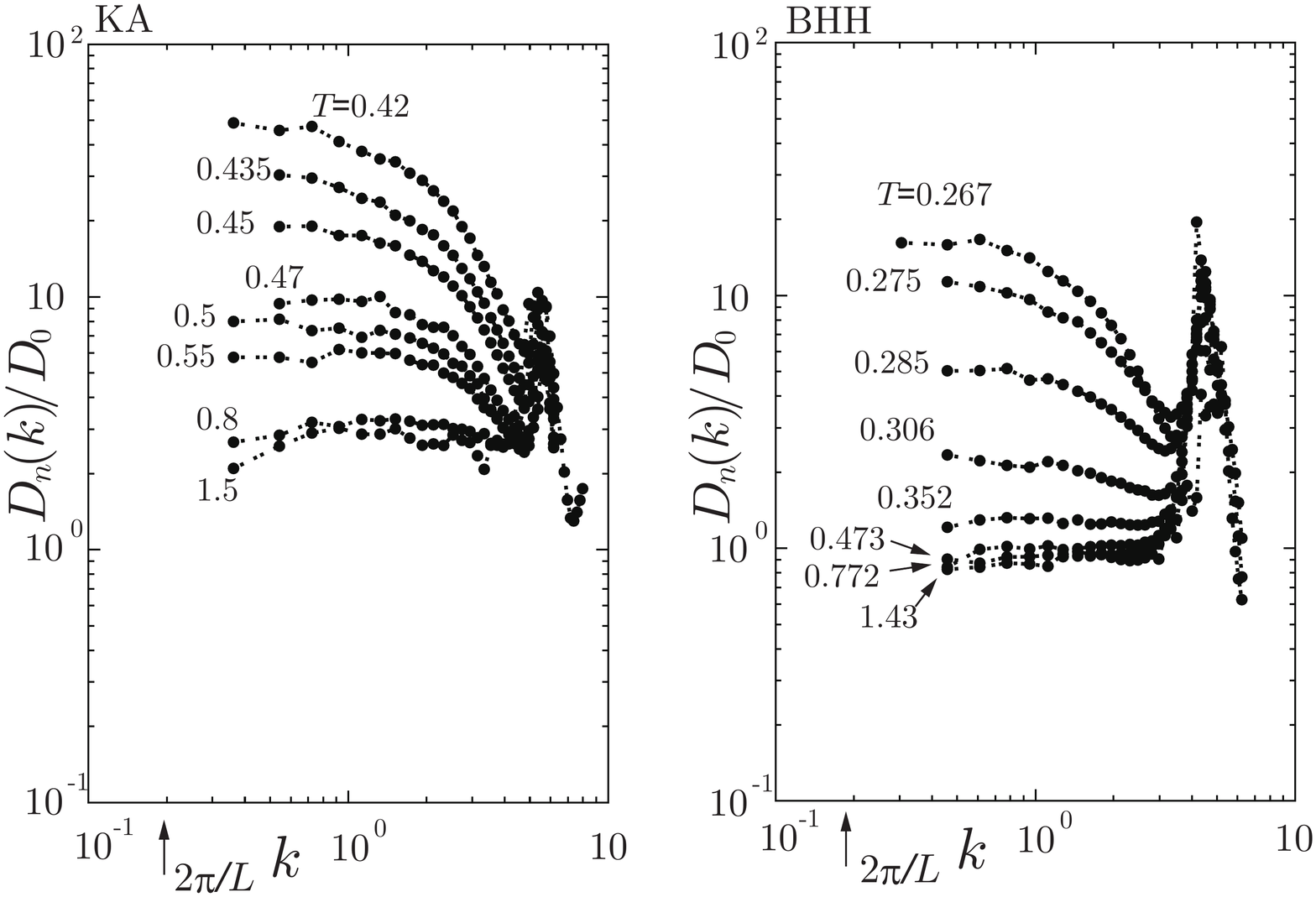}
\caption{The $k$-dependent density diffusivity $D_n(k)$ scaled by $D_0=T(2\pi\lambda\eta)^{-1}$ at various temperatures for the KA and BHH models. In higher temperature normal states, $D_n(k)$ almost shows a flat $k$ dependence. On the other hand, in lower temperature supercooled states, the deviation from the constant diffusivity is enhanced. 
}
\label{Fig2}
\end{figure}

Here we would like to emphasize that these observations for the density diffusion cannot be explained within the framework of conventional macroscopic hydrodynamics (or of the theory of simple liquids) \cite{Hansen_McdonaldB,Boon_YipB}. It is well known that conventional macroscopic hydrodynamics predicts that the scattering function spectrum of liquids consists of the Rayleigh peak and the Brillouin doublet. The Rayleigh peak is due to the heat mode, and the relaxation time of density fluctuations at long wavelengths is given by $1/D_Tk^2$, where $D_T$ is the thermal diffusivity. However, this simple picture no longer applies for (fragile) glass-forming liquids, in which a much slower structural relaxation process dominates the longitudinal transport properties; the thermal diffusivity exhibits a much weaker temperature dependence than that of the observed density diffusivity. That is, in low temperature glass-forming liquids, beacuse the heat transfer is a much faster process, the heat mode is decoupled from the structural relaxation, as well as the density diffusion.

\subsection{The growth of the cooperativity in the density exchange process and its relationship with the density diffusion}

To further investigate the cooperativity in the relaxation dynamics of density fluctuations, here we analyze the spatial correlation of the exchange events. For the calculation of the coordination number, please refer to the Appendix. Let us define 
\begin{eqnarray}
w(\mbox{\boldmath$r$},t;\Delta t)=\sum_{i=1}^{N}|\Delta z_{i}(t;\Delta t)|\delta[{\mbox{\boldmath$r$}}-{\mbox{\boldmath$r$}}_i(t)], 
\end{eqnarray}
where $\Delta z_{i}(t;\Delta t)=z_{i}(t+\Delta t)-z_{i}(t)$ is the coordination-number change over time $\Delta t$.   
The correlation function 
\begin{eqnarray}
g_w(r;\Delta t) &=& \frac{1}{N\rho}\sum_{ij}\langle |\Delta z_{i}(t;\Delta t)||\Delta z_{j}(t;\Delta t)|\delta[\mbox{\boldmath$r$}-\mbox{\boldmath$r$}_{i}(t)+\mbox{\boldmath$r$}_{j}(t)] \rangle - {\hat w}^2   
\end{eqnarray} 
measures the spatial correlation of the fluctuations of the coordination-number change or the exchange-events. 
Here, $\rho=NL^{-3}$ and ${\hat w} = N^{-1}\sum_{i=1}^{N}\langle |\Delta z_{i}(t;\Delta t)| \rangle$.  The structure factor of $w(\mbox{\boldmath$r$},t;\Delta t)$ is given as  
\begin{eqnarray}
S_w(k;\Delta t) = \frac{1}{N\rho}\langle |w_{\mbox{\boldmath$k$}}(t;\Delta t)|^2\rangle, 
\end{eqnarray} 
where $w_{\mbox{\boldmath$k$}}(t;\Delta t)=\sum_{i=1}^{N}|\Delta z_{i}(t;\Delta t)| \exp[i {\mbox{\boldmath$k$}}\cdot{\mbox{\boldmath$r$}}_i(t)]$ is the Fourier transform of $w(\mbox{\boldmath$r$},t;\Delta t)$. Note that in the definition of $w(\mbox{\boldmath$r$},t;\Delta t)$ we use the absolute value of $\Delta z_i$ instead of its true value, because $\Delta z_i$ changes very rapidly from particle to particle. Therefore, the spatial correlation of the coordiation-number change decays very fast irrespective of the degree of supercooling and thus cannot describe the correlation of the events themselves. 

For the BHH model, in Figs. 3(a) and (b),  we show $g_w(r;\Delta t)$ and $S_w(k;\Delta t)$ for $\Delta t=\tau_\alpha$ at various temperatures, respectively. Here $\tau_\alpha$ is identified as the relaxation time of the macroscopic shear stress autocorrelation. In both figures it is evident that with increase in the degree of supercooling (decreasing the temperature) the exchange process becomes more correlated. In normal states $g_w(r,\Delta t)$ exhibits the short-range correlation, resulting in an almost flat $k$-dependence of $S_w(k;\Delta t)$. On the other hand, in lower temperature supercooled states  $g_w(r;\Delta t)$ has a longer correlation length. This growth in the correlation strongly suggests that the exchange events themselves are more heterogeneous and cooperative with increasing the degree of supercooling, which is inferred as one aspect of the dynamic heterogeneity \cite{HarrowellG,Donati-Glotzer-Poole-Kob-Plimpton,Yamamoto-Onuki,Doliwa-Heuer,Laevic-Starr-Schroder-Glotzer}. 
In supercooled states the correlation length of the exchange events, $\xi_{\rm d}$, is determined from $g_w(r;\Delta t)$ or $S_w(k;\Delta t)$, which is comparable to the dynamic heterogeneity size determined by the usual four-point correlation function \cite{Laevic-Starr-Schroder-Glotzer}.  In Fig. 4, we show snapshots of the actual spatial patterns of the exchange events for the duration of the structural relaxation for the two-dimensional BHH model, which directly indicates that the exchange events are more heterogenous and correlated in supercooled states, as mentioned above.  Importantly, in supercooled states a region where larger and rather coherent displacements are found almost corresponds to a region where the correlated exchange events are found.  On the other hand, in normal states, the exchange events and the displacement fields are also significantly correlated, but they are less concentrated and more sparse than in supercooled states.  

Because the dynamic heterogeneity is now a well-established concept in glass physics, even though the correlation of the mutual particle exchange has not been investigated so far, one may think that this growth in the cooperativity in the exchange dynamics is expected without performing the above analysis. However, the relationship between the dynamic heterogeneity measured by the usual four-point correlation function method and the density diffusion is not trivial. As emphasized above, the mutual particle (density) exchange is an elementary process of the density diffusion, and thus the present observation provides clear evidence that, in normal states, the relaxation of density fluctuations takes place locally at the single particle level but in supercooled states it is increasingly cooperative and nonlocal.

\begin{figure}[htb] 
\includegraphics[width=0.8\textwidth]{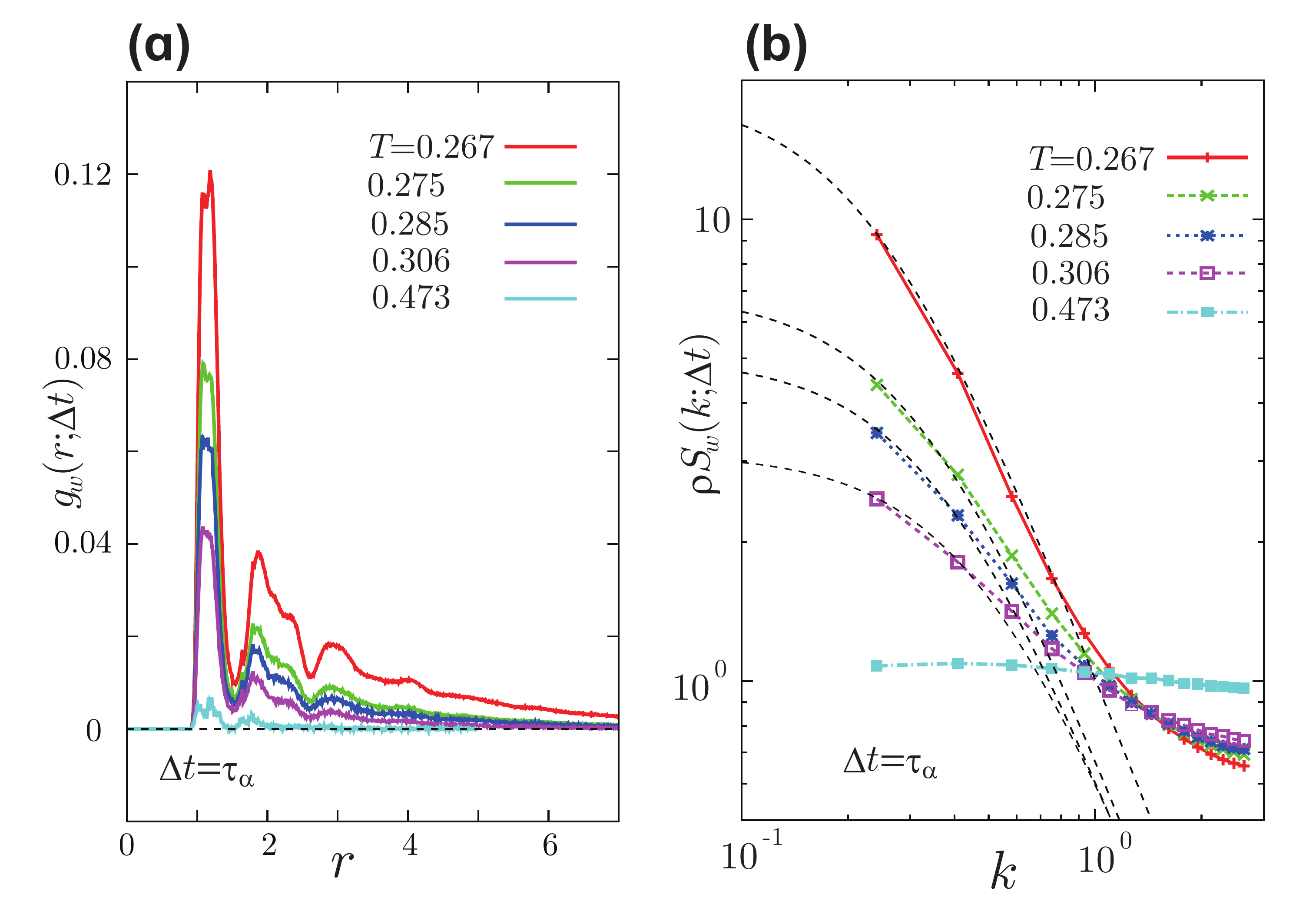}
\caption{
(a) $g_w(r;\Delta t=\tau_\alpha)$ at various temperatures for the BHH model. 
(b) $S_w(k;\Delta t=\tau_\alpha)$ at various temperatures for the BHH model. The dashed lines represent the fits ${\hat h}/(1+(\xi_{\rm 
d} k)^2)$ in supercooled states, where $\hat h$ is a numerical constant and the correlation length $\xi_{\rm d}$ corresponds to 
that determined in Fig. 5(a).   
}
\label{Fig2}
\end{figure}

\begin{figure}[htb] 
\includegraphics[width=1\textwidth]{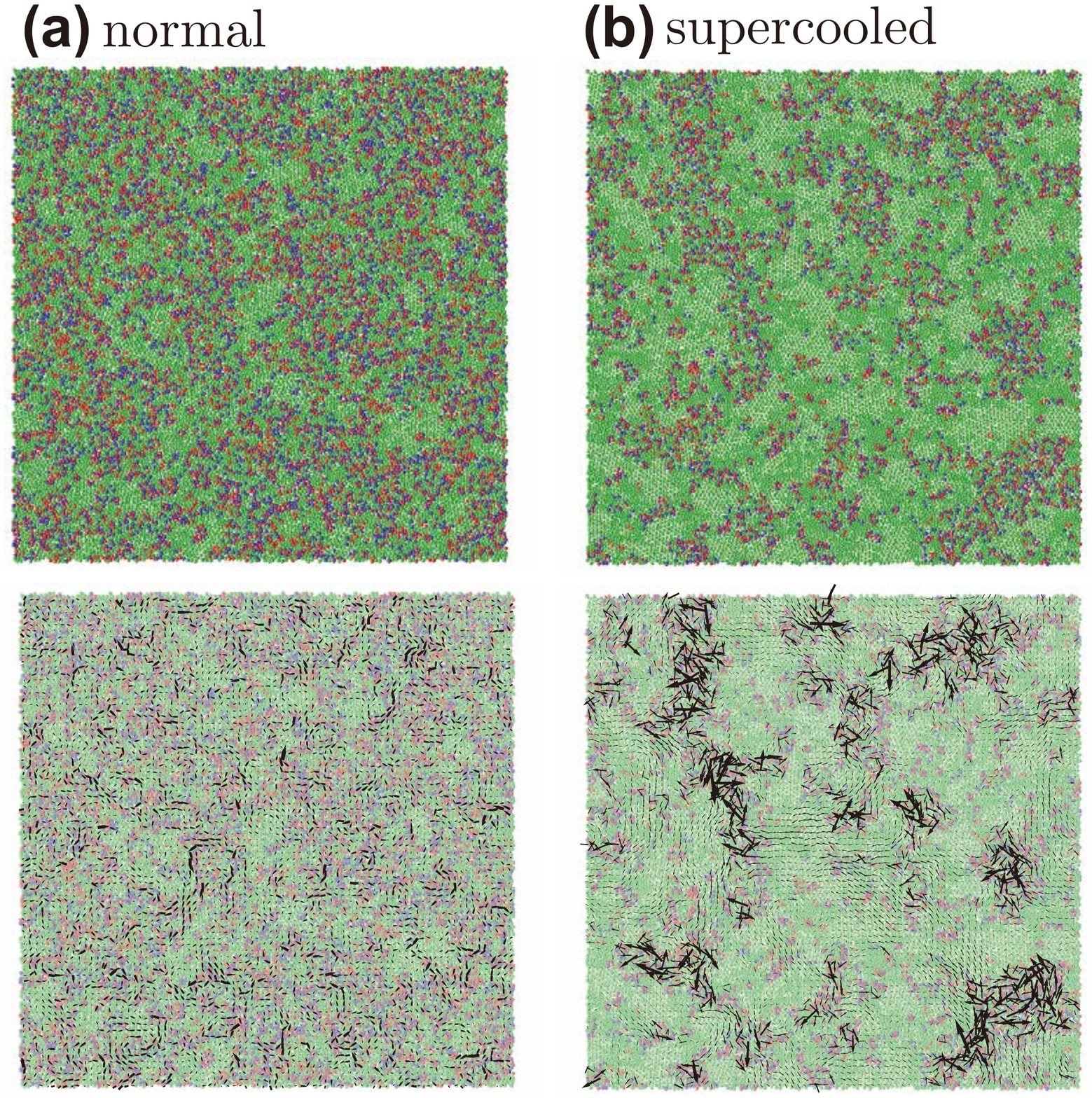}
\caption{
Upper panels: snapshots of the actual spatial patterns of the exchange events for the duration of the structural relaxation for the two-dimensional BHH model in normal (left) and supercooled (right) states. The red, blue, and green colored particles represent positive, negative, and zero coordination number change, respectively. In supercooled states, the exchange events occur in a more cooperative and heterogeneous manner. Lower panels: the displacement vector field defined by ${\mbox{\boldmath$u$}}({\mbox{\boldmath$r$}}_J;\tau_\alpha)=V_j^{-1}\sum_{i \in V_J} [{\mbox{\boldmath$r$}}_i(t+\tau_\alpha)-{\mbox{\boldmath$r$}}_i(t)]$ corresponding to the snapshots shown in the upper panels. Here,  the summation is taken over particles whose center of masses belong to a square area element $V_J$ (linear size: 1.98$\lambda_{\rm A}$) located at ${\mbox{\boldmath$r$}}_J$. In supercooled states a region where larger and rather coherent displacements are found almost corresponds to a region where the correlated exchange events are found. }
\label{Fig4}
\end{figure}

From these observations, we immediately draw the following physical picture for the density diffusion.  In normal liquid states at higher temperatures, the dynamics at the single particle scale ($\sim \lambda$) dominate the density transport. Namely, the density diffusion is determined by less correlated (or correlated only for distances of the particle size $\lambda$) exchange events for the time duration of $\tau_\alpha$, resulting in the following expression of the density diffusivity: 
\begin{eqnarray}
D_c \sim \frac{T}{2\pi \eta\lambda} \sim \frac{\lambda^2}{\tau_\alpha}, ~~~~({\rm in~~normal~~liquid~~states})  \label{diffusivity_normal}
\end{eqnarray}
where we make use of the Maxwell relation $\eta\cong E\tau_\alpha$, with $E(\sim T/\lambda^3)$ being the shear modulus. However, as the degree of supercooling is increased, the single-particle scale dynamics are surpassed by the cooperative dynamics associated with the growing dynamic correlation: In supercooled states, the density diffusion occurs via the cooperative exchange events over distances of $\xi_{\rm d}$ and times of $\tau_\alpha$, which are the characteristic length and time scales for the diffusion in supercooled states, respectively ($\xi_{\rm d}$ is regarded as a unit size). 
Consequently, we obtain \cite{FurukawaG3,FurukawaG4,FurukawaG5} 
\begin{eqnarray}
D_{c}\sim \frac{\xi^2_{\rm d}}{\tau_\alpha}
\biggl(\gg \frac{\lambda^2}{\tau_\alpha}\biggr).~~~~({\rm in~~supercooled~~liquid~~states}) \label{diffusivity_supercooled}
\end{eqnarray}

To check the validity of the above argument, we analyze the scaling properties of the $k$-dependent density-diffusion coefficient $D_n(k)$. In Fig. 2 we show $D_n(k)$ scaled by $D_0=T(2\pi \lambda\eta)^{-1}( \sim \lambda^2 \tau_\alpha^{-1})$. In these plots the particle size $\lambda$ is set to the unit length. At higher temperatures (in normal liquid states), $D_n(k)$ shows a constant diffusivity as $D_n(k) \sim T(\lambda\eta)^{-1} (\sim \lambda^2\tau_\alpha^{-1})$ for length-scales larger than the single particle size. However, as the degree of supercooling increases, the deviation from this constant diffusivity is enhanced. As described above, the collective density relaxation in supercooled states can be viewed as a consequence of the cooperative exchange events (and probably rather coherent motions of the particles), which leads to Eq. (\ref{diffusivity_supercooled}). In Fig. 5(b), we show the diffusion coefficient $D_n(k)$ scaled by $D_1= \xi^2_{\rm d}(2 \pi \tau_\alpha)^{-1}$ in supercooled states as a function of the scaled wavenumber $k\xi_{\rm d}$. We find that $D_n(k)/D_1$ nearly falls onto a single master curve. In Ref. \cite{FurukawaG3} we showed similar plots with a different definition of the correlation length, whose scaling is worse than the  present one. For $k\xi_{\rm d} < 1$, the slowly relaxing density fluctuations obey the diffusion equation with a diffusion coefficient $D_n(k)\sim \xi^2_{\rm d}\tau_\alpha^{-1}$. On the other hand, for $k\xi_{\rm d}> 1$ the relaxation time is nearly equal to the $\alpha$-relaxation time, $\tau_n(k)\sim \tau_\alpha\propto \eta/T$; thus, density fluctuations survive for a time-scale of $\tau_n(k)\sim \tau_\alpha$.  These observations strongly support the scenario of cooperative diffusion. 
\begin{figure}[htb] 
\includegraphics[width=1.1\textwidth]{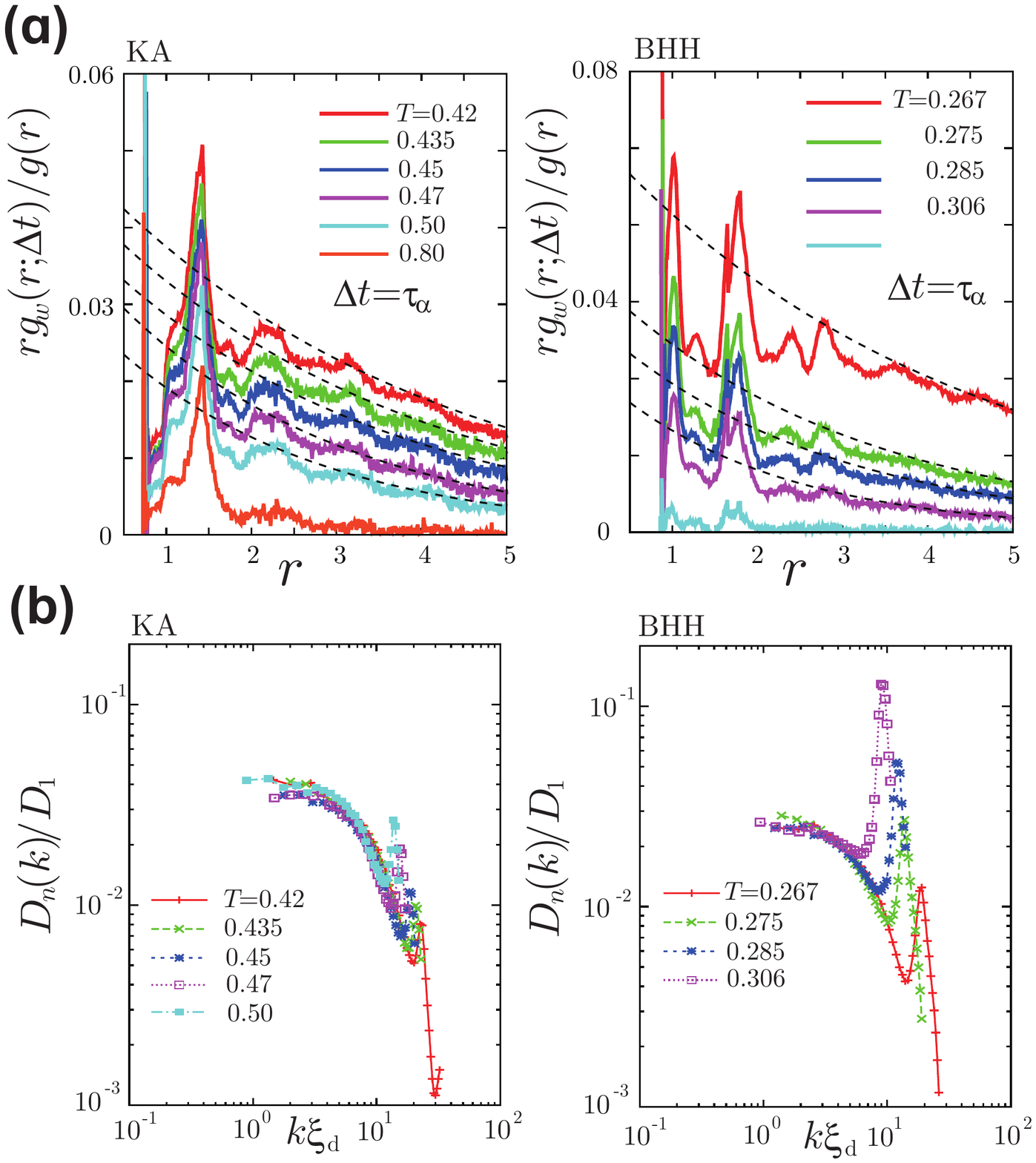}
\caption{(a) $g_w(r;\Delta t=\tau_\alpha)$ scaled by $r^{-1}g(r)$ for the KA and BHH models, where $g(r) = ({N\rho})^{-1}\sum_{i}\langle \delta(\mbox{\boldmath$r$}-\mbox{\boldmath$r$}_{i})\rangle$ is the pair correlation function.  The dashed lines are the fits to the Ornstein Zernike form,  $g_w(r;\Delta t=\tau_\alpha)/g(r) \propto r^{-1} \exp(-r/\xi_{d})$ for larger $r$.  
(b) The nonlocal diffusivity $D_n(k)$ scaled by $D_1=\xi_{\rm d}^2 (2\pi\tau_\alpha)^{-1}$ as a function of the scaled wavenumber $k\xi_{\rm d}$ in supercooled states for the KA and BHH models. $D_n(k)/D_1$ nearly falls onto a single master curve. 
}
\label{Fig5}
\end{figure}

\subsection{Hydrodynamic perspective}

The expressions of the density diffusivity, Eqs. (\ref{diffusivity_normal}) and (\ref{diffusivity_supercooled}) in normal and supercooled states, respectively, can also be obtained by the following simple hydrodynamic arguments: Hereafter, we refer to the slowly relaxing density as $\hat\rho$. We assume that the restoring force associated with density fluctuations balances the friction force as 
\begin{eqnarray}
0\cong -\zeta{\mbox{\boldmath$v$}}_D-\frac{1}{K_0}\nabla \frac{\hat \rho}{\hat \rho_0}, \label{force_balance}
\end{eqnarray}
where $K_0$ is the compressibility in the long-wavelength regime, $\hat\rho_0$ is the average value of $\hat\rho$, and ${\mbox{\boldmath$v$}}_D$ represents the diffusive velocity. Therefore, $\hat\rho{\mbox{\boldmath$v$}}_D$ provides the ``diffusive current''. Recall that the static two-body density correlations remain almost unchanged during the vitrification, so that the restoring force and $K_0$ should be insensitive to the degree of supercooling and thus to $\xi_{\rm d}$. On the other hand, the friction cofficient $\zeta$ can directly reflect the cooperative effect. In other words, we implicitly suppose that the dynamic correlation length for the density diffusion $\xi_{\rm d}$ cannot be identified through the static two-body density correlator or through the free energy functional.  

In normal states, the substantial origin of the density diffusion is the less correlated particle diffusion caused by the mutual particle exchange events. For this, by applying Stokes law to the particle, the friction coefficient (per unit volume) is evaluated as 
\begin{eqnarray}
\zeta\sim 2\pi \eta\lambda \times \lambda^{-3} ~~~~({\rm normal~~liquid~~states})  \label{friction_normal}
\end{eqnarray} 
Because $\hat\rho$ evolves by the continuity equation,  with Eqs. (\ref{force_balance}) and,  (\ref{friction_normal}),  
\begin{eqnarray}  
\frac{\partial{\hat\rho}}{\partial t} \cong -\hat\rho_0\nabla\cdot {{\mbox{\boldmath$v$}}}_D \sim \frac{1}{K_0 \zeta} \nabla^2 \hat \rho, \label{continuity}
\end{eqnarray}
resulting in 
\begin{eqnarray}
D_c\sim \frac{1}{K_0\zeta} \sim \frac{\lambda^2}{\tau_\alpha},  
\end{eqnarray}
which is nothing but Eq. (\ref{diffusivity_normal}), where the relationships $K_0E \sim 1$ and $\eta\sim E \tau_\alpha$ are used. 

On the other hand, in supercooled states, comparing Eqs. (\ref{diffusivity_normal}) and (\ref{diffusivity_supercooled}), the density diffusivity is enhanced by $(\xi_{\rm d}/\lambda)^2$, which indicates that the friction coefficient $\zeta$ is reduced by the same factor. This enhancement in $D_c$ and the reduction in $\zeta$ are expected to result from the increasing cooperativity: we assume that the correlated exchange events produce rather coherent particle motions on length scales of $\xi_{\rm d}$ and that such coherent motions are the physical substance of the diffusive current.  For this assumption, $\zeta$ can be identified with the friction coefficient for the transiently correlated fluctuations with a size of  $\xi_{\rm d}$, leading to  
\begin{eqnarray}
\zeta\sim \eta\xi_{\rm d} \times \xi^{-3}_{\rm d},  ~~~~({\rm supercooled~~liquid~~states}) \label{friction_supercooled}
\end{eqnarray}
and 
\begin{eqnarray}
D_c\sim \frac{1}{K_0\zeta} \sim \frac{\xi_{\rm d}^2}{\tau_\alpha}. 
\end{eqnarray}
To verify the present hydrodynamic argument for the density diffusivity in supercooled states, some further intensive study is required, which will be the subject of a future study. 

\section{Breakdown of the Stokes Einstein-relation}

We have so far argued that the change in the degree of cooperativity can be manifested in the transport crossover in the density diffusion. The coherent scattering function at wavenumber $k$ measures the relaxation of density fluctuations at size $\sim 1/k$, and thus this crossover phenomenon should directly reflect the emergence and growth of cooperative motions. In a series of our previous studies \cite{FurukawaG1,FurukawaG2,FurukawaG3}, we have discussed the problem involving the breakdown of the Stokes-Einstein (SE) relation via the collective density-diffusion instead of the single-particle diffusion. However, the breakdown of the SE relation is primarily about the self-diffusion of a tagged particle \cite{Fujara,Cicerone-Ediger,Angell-Ngai-McKenna-Mcmillan-Martin}. Therefore, we have implicitly assumed that the collective-diffusion is governed by the same dynamical process as in self-diffusion. Here we provide supporting evidence for this assumption.

 In Fig. 6, for both the KA and BHH models, we show the temperature dependence of the self- and collective-diffusion coefficients, $D_s$ and $D_c$, respectively. The (total) self-diffusion coefficient, $D_s$, is determined by $D_s= \lim_{\Delta t\rightarrow \infty}{\langle |\Delta r_i|^2\rangle}/{6\Delta t}$, where $\langle |\Delta r_i|^2\rangle$ is the mean-square displacement of the particles for the time duration, $\Delta t$, and an average is taken over all of the particles. On the other hand, as described above, $D_c$ is formally identified as $D_c=\lim_{k\rightarrow 0}D_n(k)$, where $D_n(k)$ is the $k$-dependent density-diffusion coefficient, which is shown in Fig. 2. Figure 6 shows that $D_c$ and $D_s$ have almost the same temperature dependence for the temperature range of the present study. Specifically, in the BHH model $D_c$ and $D_s$ almost collapse onto a single curve. This strong correlation between $D_c$ and $D_s$ supports our assumption and can be interpreted as follows: in normal states at higher temperatures, the particle dynamics are less-cooperative, resulting in the SE relation holding well. Such single-particle scale dynamics dominate the hydrodynamic transport; thus, $D_s(\sim \lambda^2/\tau_\alpha)$ determines $D_c$. However, in supercooled states at lower temperatures, the deviation from the SE relation is significant, which suggests that the cooperative dynamics surpass the particle dynamics; thus, $D_c(\sim \xi_{\rm d}^2/\tau_\alpha)$ determines $D_s$, resulting in the SE relation being violated by factor $(\xi_{\rm d}/\lambda)^2$. In this sense, the breakdown of the SE relation is not an anomaly of the single-particle or local dynamics, but instead reflects the emergence and growth of the dynamical cooperativity.  

In the KA model, a slight difference between $D_s$ and $D_c$ ($D_c> D_s$) can be found. We naively interpret this difference as follows: In the BHH model, local volume change is not allowed due to the additive interaction. Accordingly, the local conservation law strictly holds. Therefore, both $D_s$ and $D_c$ reflect the same dynamical process. That is, the particle and density diffusion proceed only via the mutual particle exchange process. On the other hand, in the KA model, because of the non-additive cross interaction diameter, there are kinetic paths where a significant local volume change (or the relaxation of density fluctuations) can take place without a mutual particle exchange. This may explain why $D_c>D_s$. However, these two dynamics should still be governed mainly by the exchange process as suggested by the almost identical temperature dependences of $D_s$ and $D_c$. 

\begin{figure}[h]
\includegraphics[width=0.8\textwidth]{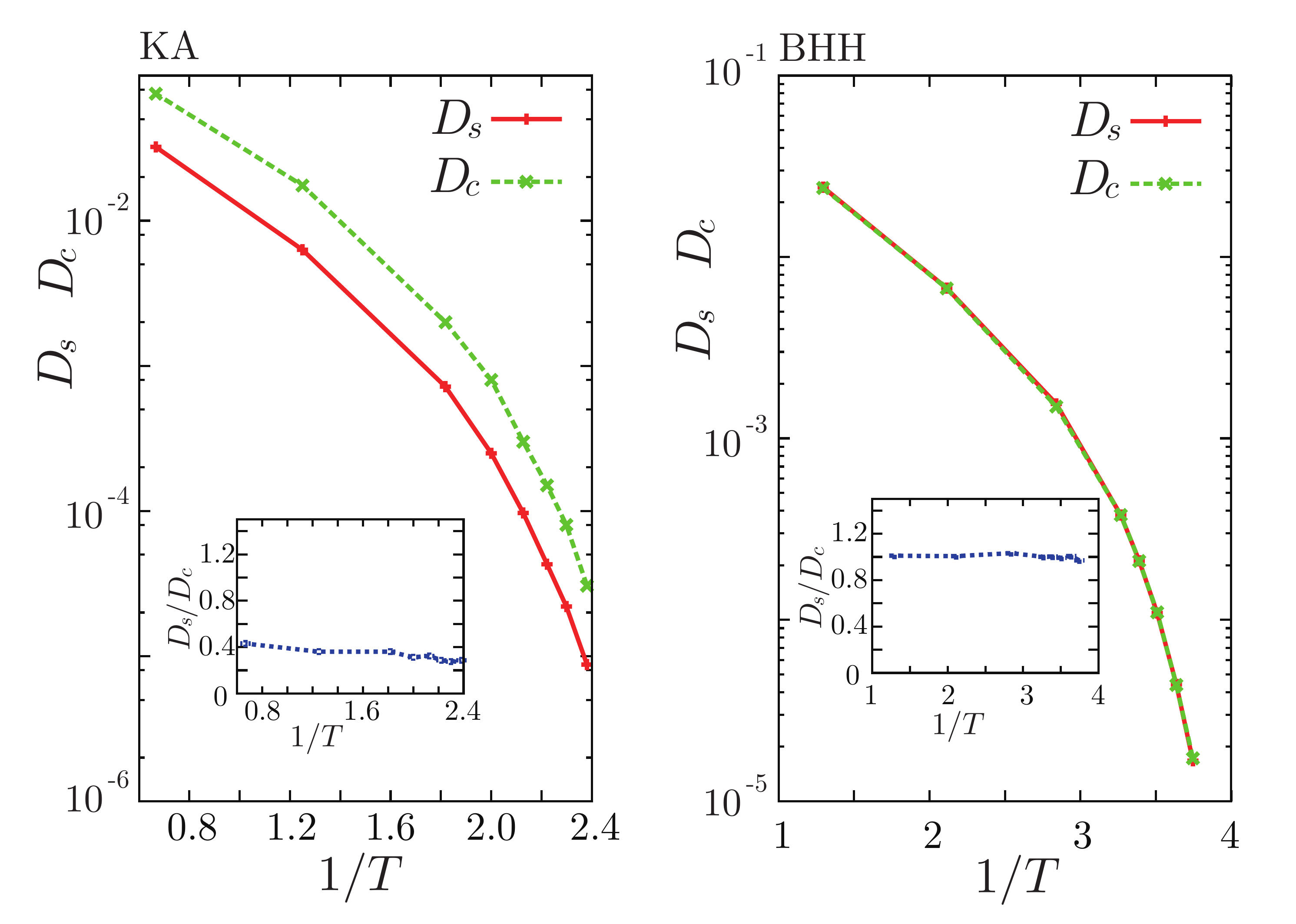}
\caption{
The temperature dependences of the self- and collective-diffusion coefficients of the KA and BHH models. In the BHH model, $D_c$ and $D_s$ almost collapse onto a single curve.  The insets show the ratio $D_s/D_c$. 
}
\label{Fig6}
\end{figure}

\section{Discussion and Remarks}

In this paper, contrary to the conventional view, for fragile glass-forming liquids, we have discussed that the emergence and growth of the cooperativity associated with vitrification can be captured by the two-body dynamic density correlator.  It is emphasized that, so far, very little attention has been paid to the potential of the density diffusion to investigate the cooperative effects. Although the main argument is similar to that in Refs. \cite{FurukawaG1,FurukawaG2,FurukawaG3} by the present author, here we provided further evidence for the phenomenology of the density diffusivity and the breakdown of the SE relation. In normal liquid states at higher temperatures the density diffusion is dominated by less correlated (or correlated only for distances of the particle size $\lambda$) particle exchange dynamics for the time duration of $\tau_\alpha$, resulting in $D_c \sim {\lambda^2}{\tau_\alpha}^{-1}$, while in supercooled states the density exchange occurs cooperatively over distances of $\xi_{\rm d}$ and times of $\tau_\alpha$, leading to $D_{c}\sim {\xi^2_{\rm d}}{\tau_\alpha}^{-1}$. 

We note the following point. In normal states, $D_n(k)$ do not show significant differences between microscopic ($k\sim 2\pi\lambda^{-1}$) and macroscopic ($k\sim 0$) scales, that is, the macroscopic hydrodynamic theory can be applied over quite a wide range of spatial (and temporal) scales. This is because there is no important characteristic (static or dynamic) length scale beyond the particle size, and the single-particle scale dynamics determine the hydrodynamic transport properties. Thus, roughly speaking, hydrodynamic fluctuations (here, density fluctuations) relax in an uncorrelated manner and follow the particle motions: $D_s$ determines $D_c$. On the other hand, this is not the case in supercooled liquids. The density diffusion is determined by the cooperative density exchange dynamics, to which smaller scale ($<\xi_{\rm d}$) fluctuations and thus the single particle dynamics are subordinated: $D_c$ determines $D_s$.

In our argument on the density diffusivity, we assumed that the cooperative length scale $\xi_{\rm d}$ is responsible not for the thermodynamic force but for the friction coefficient, whence such the cooperative effects may have a dynamic origin. In glass physics, many efforts have been devoted to elucidating the origin of the cooperativity or growing length scale observed in supercooled liquids. Recent theoretical attempts have been put forward, particularly from thermodynamic perspectives. The most representative one is the random-first-order-transition theory \cite{Kirkpatrick-Thirumalai-Wolynes,Lubchenko-Wolyness}, which was developed based on the Adam-Gibbs theory \cite{Adam_Gibbs} with the concept of the spin-glass physics. On the other hand, there are different perspectives based on the nonsingular density fluctuations. Among them, the two major representatives, the free volume theory \cite{Cohen-Turnbull,Turnbull-Cohen} and the mode coupling theory \cite{GotzeB}, have been intensively studied so far. However, it turns out that they hardly include any notions of the cooperativity or growing length scale. This may be one reason why (still unknown) thermodynamic singularities are expected as the origin of the growing length scale and eventually the glass transition. However, is this truly the case? In Ref. \cite{FurukawaG5}, another perspective was given also based on nonsingular density fluctuations. In a fragile glass-forming liquid near the glass transition point, a small change in the macroscopic average density determines the macroscopic glass transition. Concomitantly, the density itself fluctuates in space. Therefore, one may imagine that even a slight change in the local density should control the local glassy nature. Based on this perspective, the present author constructed a theoretical model, in which the concept of a growing length scale can be naturally introduced without invoking thermodynamic anomalies. We are currently investigating the predictability and verifiability of this model, which will be discussed elsewhere.  

\section*{Acknowledgement}
This work was supported by KAKENHI (Grant No. 26103507, 25000002, and 20508139), the JSPS Core-to-Core Program ``International research network for non-equilibrium dynamics of soft matter'', and the special fund of Institute of Industrial Science, The University of Tokyo. 
\appendix
\section{Simulation Models}

In this study, we used two simple and popular model fragile glass-forming binary mixtures: the Kob-Andersen (KA) \cite{Kob-Andersen} and the Bernu-Hiwatari-Hansen (BHH) soft-sphere \cite{Bernu-Hiwatari-Hansen} models. These models were simulated using velocity Verlet algorithms in the NVE ensemble \cite{RapaportB}. Here, we describe the details of these model systems. 

{\it The KA model.---} The Kob-Andersen (KA) model \cite{Kob-Andersen} is a binary mixture, which is composed of large ($A$) and small ($B$) particles of equal masses, $m_A=m_B=m$. The interaction potential is given by   
\begin{eqnarray}
U_{\mu\nu}^{\rm KA}(r)=4\epsilon_{\mu\nu}\biggl[ \biggl(\frac{\lambda_{\mu\nu}}{r}\biggr)^{12}-\biggl(\frac{\lambda_{\mu\nu}}{r}\biggr)^{6}\biggr] - U_{\mu\nu}^0,  
\end{eqnarray}
where $\mu,\nu= A,B$, $\epsilon_{AB}=1.5\epsilon_{AA}$, $\epsilon_{BB}=0.5\epsilon_{AA}$, $\lambda_{AB}=0.8 \lambda_{AA}$, $\lambda_{BB}=0.88 \lambda_{AA}$ and $r$ is the distance between two particles. The potential is truncated at $r=2.5\lambda_{\mu\nu}$ and $U_{\mu\nu}^0$ is chosen to satisfy  $U_{\mu\nu}^{\rm KA}(2.5\lambda_{\mu\nu})=0$. The temperature $T$ was measured in units of $\epsilon_{AA}/k_B$.  
We held the particle number density constant at a value of $N/V =1.2/\lambda_{A}^{3}$, where $N=N_A+N_B=36000$ with $N_A/N_B=4$, and $V$ is the system volume. The space and time units were $\lambda_{AA}$ and $(m \lambda_{AA}^{2}/\epsilon_{AA})^{1/2}$, respectively. Then, the linear dimension of the system was $L=31.07$. 

{\it The BHH model.---} 
The Bernu-Hiwatari-Hansen model \cite{Bernu-Hiwatari-Hansen} is a binary mixture of large ($A$) and small ($B$) particles interacting via the soft-core potentials
\begin{eqnarray}
U_{\mu\nu}^{\rm BHH}(r)=\epsilon \biggl(\frac{\lambda_{\mu\nu}}{r}\biggr)^{12}, 
\end{eqnarray}
where $\mu,\nu= A,B$, $\lambda_{\mu\nu}=(\lambda_{\mu}+\lambda_{\nu})/2$, $\lambda_{\mu}$ is the particle size, and $r$ is the distance between two particles. The mass is $m_{B}/m_{A}=2$, and the size ratio is $\lambda_{B}/\lambda_{A}=1.4$ (1.2) in 2D (3D). The units for the length and time are $\lambda_A$ and $({m_{A}\lambda_{A}^{2}/\epsilon})^{1/2}$, respectively. The total number of particles was $N=N_{A}+N_{B}=20000$ (40000) in 2D (3D). Here, $N_A/N_B=1$. The temperature $T$ was measured in units of $\epsilon/k_{\rm B}$. The fixed particle number density and the linear dimension of the system were $N/V =0.8/\lambda_{A}^{d}$ and $L=158.11$ (36.84) in 2D (3D), respectively. 

In the present binary mixtures, the coordination number of the $i$-th particle of species $\mu$ at time $t$, $z_i^\mu(t)$, is defined as the total number of particles satisfying $|\mbox{\boldmath$r$}_i^\mu(t)-\mbox{\boldmath$r$}_j^\nu(t)|<r_{\rm min}^{\mu\nu}$. At $r=r_{\rm min}^{\mu\nu}$, the radial distribution function $g_{\mu\nu}(r)$ has its first minimum. In the main text, we simply denote $z_i^{\mu}(t)$ as $z_i(t)$.

\end{document}